\documentstyle[12pt]{article}
\input{psfig}

\def\a{\alpha}

\def\g{\gamma}

\def\({\left(}
\def\){\right)}

\def\citenum#1{{\def\@cite##1##2{##1}\cite{#1}}}
\def\citea#1{\@cite{#1}{}}

\def\beq{\begin{equation}}
\def\eeq{\end{equation}}
\def\bea{\begin{eqnarray}}
\def\eea{\end{eqnarray}}

\def\bbbz{{\mathchoice {\hbox{$\sf\textstyle Z\kern-0.4em Z$}}
{\hbox{$\sf\textstyle Z\kern-0.4em Z$}}
{\hbox{$\sf\scriptstyle Z\kern-0.3em Z$}}
{\hbox{$\sf\scriptscriptstyle Z\kern-0.2em Z$}}}}
\relax

\begin{document}
\begin{flushright}
{TAUP 2307-95 \\
CBPF NF-085/95 \\
UCI 95-34 \\
December 1995}
\end{flushright}
\vskip 1 true cm
\begin{center}
\Large {\bf A Screened BFKL Interpretation of $F_2$ in \\
the Exceedingly Small x Limit}
\end{center}
\vskip 8pt
\begin{center}
E. GOTSMAN$^{a,b,}$\footnote{Email: gotsman@ccsg.tau.ac.il},
E. M. LEVIN$^{c,d,}$\footnote{Email: levin@lafex.cbpf.br}
 and U. MAOR$^{a,c,}$\footnote{Email: maor@vm.tau.ac.il}
\vskip 4pt
{\sl $^{a}$ School of Physics and Astronomy\\
Raymond and Beverly Sackler Faculty of Exact Science\\
Tel Aviv University,
Tel  Aviv, 69978, Israel}\\
\vskip 3pt
{\sl $^{b}$ Physics Department, University of California\\
Irvine, CA 92717, USA}\\
\vskip 3pt
{\sl $^{c}$ LAFEX, Centro Brasileiro de Pesquisas F\'\i sicas (CNPq)\\
Rua Dr. Xavier Sigaud 150, 22290 - 180, Rio de Janiero, RJ, Brasil}
\vskip 3pt
{\sl $^{d}$ Theory Department,
St. Petersburg Nuclear Physics Institute\\
188350, St. Petersburg, Gatchina, Russia}
\end{center}
\vskip 5pt

\begin{abstract}
In this letter we show that the behaviour of $F_{2}$, at very
small $x_B$,
agrees with the behaviour expected from the BFKL evolution
equation, when  screening corrections are included.
We obtain a description which is consistent with the data, however, we require
 the screening corrections
 to be relatively large (about a quarter of the total DIS cross
section). The relation between the screening corrections
and the diffractive DIS cross section is discussed.
\end{abstract}

\newpage
In this letter we discuss the
dependence of
$F_{2}$, the proton structure function, on W, the $\gamma^* p$ c.m.
energy, at very small $x_{B}$. We are motivated by the recently published
data taken at HERA by the ZEUS \cite{ZEUS} and H1 \cite{H1}
collaborations which are shown in Fig.1.
Our goal is to extract new information from
the experimental data  on the deep
inelastic scattering (DIS) process, in the region of very small
$x_{B}$.  We will show
that the BFKL evolution equation (the BFKL Pomeron)
\cite{BFKL},  including screening (shadowing)
corrections\cite{GLR},
provides a good reproduction of
the observed data.
\newline
\par
We list first the main qualitative properties of the behaviour of
$F_2(W,Q^2)$, as observed by the two experimental groups at HERA and
shown in Fig.1.
\\
1) For W values below 130-150 GeV, the measured data points cluster
in a narrow linear band. i.e. $F_2(W,Q^2)$ is approximately linear
in W and has a weak dependence on $Q^2$.
\\
2) For higher values of W, $F_2(W,Q^2)$ is dependent on $Q^2$.
The high $Q^2$ data  differs
from the lower
$Q^2$ data which seems to reach a local plateau, resulting in a $F_2$
which is almost  constant as a function of W.
\newline
\par
The features of the data at
low W  are
compatible
with a dominance of the BFKL Pomeron\cite{BFKL} in the small
$x_{B}$ domain of $F_2(x_B,Q^2)$. This is readily seen when we write
\cite{CCH} the BFKL generated structure function
\beq
F^{BFKL}_2(x_B,Q^2)\,\,=\,\,\Sigma_f e^2_f\,\cdot\, \frac{11 \pi^2
\a_s(Q^2_0)}
{32 \sqrt{2}}\,
\cdot\,\frac{G_0}{\sqrt{28 N_c \a_s(Q^2_0) \zeta(3)}}\,
\eeq
$$\,\cdot\,
\sqrt{\frac{Q^2}{Q^2_0}}\,\cdot\,\frac{1}{\sqrt{\ln \frac{1}{x_B}}}\,
\cdot\,
(\frac{1}{x_B})^{\omega_0}\,\cdot\,e^{- \,\frac{\pi(\ln\frac{Q^2}{Q^2_0})
^2}{
56 N_c \a_s(Q^2_0)\zeta(3) \ln\frac{1}{x_B}}}
$$
where $G_0$  denotes the unknown normalization of the gluon
distribution
at  $x_B \,\sim\,1$. The value of $\omega_0$ is given by\cite{BFKL}
$\omega_0\,\,=\,\,\frac{N_c \a_s}{\pi} \,\,4\,\ln 2 \,\,\,. $
In the following we  assume that
$\omega_0$ = 0.5, which corresponds
to a  resonable value of $\a_s$.
At small values of $x_B$,  $W^2 \,=
\frac{Q^2}{x_B}$, so we can rewrite Eq.(1) in the form
\beq
F_2(x_B,Q^2) \,\,\propto\,\,\frac{W}{\sqrt{\ln\frac{W}{Q}}}
\,\cdot\,e^{- \,\frac{\pi(\ln\frac{Q}{Q_0})^2}{
28 N_c \a_s(Q^2_0)\zeta(3) \ln\frac{W}{Q}}}
\eeq
The above expression reproduces the qualitative
$F_2$ features listed for the lower W, but fails to reproduce the
required high energy characteristics.
 To improve the ``theoretical'' behaviour of $F_2$ at higher
values of W, we introduce the
shadowing correction\cite{GLR}, illustrated in Fig.2.
The DIS structure function can be represented as
\beq
F_2(x_B,Q^2)\,\,=\,\,F^{BFKL}_2 (x_B,Q^2)\,\,+\,\,\Delta F_2(x_B,Q^2)
\eeq
where $\Delta F_2$ represents the changes in the BFKL structure
function which result from  screening.
\newline
\par
We shall elaborate  on the details of the screened diagram
calculation, shown in Fig.2,
later. Our discussion is based on the main results of Ref.\cite{BLW}
which are summarized as follows: The dominant
contribution of interest comes from
the exchange of two BFKL ladders (Pomerons), while the upper blob
of Fig.2 is suitably given by
the GLAP\cite{GLAP} DIS
structure function. The integration over $x_P$ (see
notations in Fig.2) results in the  contribution
\beq
\Delta F_2 (x_B,Q^2)\,\,=
\,\,-\,\, \Sigma_f e^2_f\,\cdot\,
 \frac{11 \pi^2 \a_s(Q^2_0)}{32 \sqrt{2}}\,\cdot\,\{ \frac{Q^2_0}{Q^2}
\}^{-\frac{\a_s}{2 \omega_0}}
\eeq
$$
\,\cdot\,\frac{G_0}{\sqrt{28 N_c \a_s(Q^2_0) \zeta(3)}}\,\,\cdot\,\gamma
\,\cdot \, \frac{\ln \frac{Q^2}{Q^2_0}}{\ln(\frac{1}{x_B})}\,\cdot\,
(\frac{1}{x_B})^{2 \omega_0}
$$
where we have absorbed all
nonperturbative QCD contributions in the
phenomenological triple "ladder" vertex $ \gamma $.
The minus sign in Eq.(4)
reflects the shadowing origin of this contribution.
Eq.(4) can also be derived
from the calculation of the diffraction dissociation cross section using
the AGK cutting rules \cite{AGK}, which lead to the relation
\cite{WULE}
\beq
\Delta F_2 \,\,=\,\,-\,\,F^D_2
\eeq
We  note that $F^D_2$, in the above equation, is related
to the total integrated diffractive DIS cross section.
Namely, to the  two DIS single diffraction channels, as well as the
DIS double and central diffraction. It has recently been suggested
that these are quite large\cite{NIK}\cite{GLM2}.
The restriction implied by Eq.(5) hinders  our ability to reconstruct
 ``theoretically'',
 the experimental high energy behaviour of $F_2(W,Q^2)$.
The BFKL approach, which qualitatively reproduces the low
energy features, fails to do so at high energies. We conclude from this
that the necessary  corrections,
and therefore the diffractive component, must be quite large.
As we shall see, when discussing the results of our calculation,
we require that $\frac{F_2^D}{F_2} \geq 0.25$.
This requirement is not in contradiction to the meagre DIS
experimental information presently available, allowing one to check
 $\frac{F_2^D}{F_2}$.
 Both ZEUS\cite{ZEUSG}
and H1\cite{H1G}
collaborations
find a sizeable diffractive component
in their $Q^2 \simeq 0$ photoproduction studies. In DIS, only single
diffraction at the $\g^{*}$ vertex has been
measured\cite{ZEUSD}\cite{H1D}, and the ratio of the measured diffraction
to the total DIS cross section is about 0.15. Estimates of the non
measured diffractive channels are model dependent.
Irrespective of our detailed estimate of the unmeasured channels, the
overall ratio obtained, is sufficiently large to justify our approach.
\newline
\par
As stated, we take $\omega_0 = 0.5$. Seemingly, with value of
$\omega_0$ we reach the unitarity limit\cite{GLR}, and expect the
diffractive channel to vanish. As s-channel unitarity is not built
into this formalism,  we can take  $\omega_0$
values even larger  than 0.5,
provided that unitarity corrections, such as screening,
are incorporated in the calculation.
 Absorbing  all the unknown factors in a new phenomenological constant
$\tilde G_0$ we obtain
\beq
F_2(x_B,Q^2)\,\,=\,\,\tilde G_0 \,\cdot\,\{ \frac{W}{\sqrt{\ln \frac{W}
{Q}}}
\,\cdot\,
exp[\,\, - \,\,\frac{\pi(\ln\frac{Q}{Q_0})^2}{
28 N_c \a_s(Q^2_0)\zeta(3) \ln\frac{W}{Q}} \,\,]
\eeq
$$
\,\,-\,\,W^2\,\cdot\,\frac{\gamma\,\,\ln\frac{Q}{Q_0}}
{\ln\frac{W}{Q}}\,\cdot \{\,\frac{Q^2_0}{Q^2} \,\}^{(
1 - \frac{1}{4\ln2})} \,\}\,+\,\,C
$$
We have  added a (small) constant C to account for the remnant
non BFKL contributions.
\newline
\par
We  now turn to a more detailed discussion of the diagram shown in
Fig.2. Our motivation is three fold:
\\
1) We wish to better comprehend the complicated calculation of Bartels,
Lotter and Wuesthoff\cite{BLW} and its consequences.
\\
2) We need to clarify how trustworthy  our perturbative QCD
calculation is.
\\
3) We need to adjust the results to the relevant kinematic domain at HERA.
\newline
\par
We use the
expression given for our diagram in Ref.\cite{WULE}
\beq
\Delta F_2 =-\gamma \int^{\ln\frac{1}{x_B}}_0
 d \ln \frac{1}{x_P} \int^{Q^2}_{Q^2_0} \frac{d k^2}{k^4}
F^{GLAP}_2(\frac{x_B}{x_P},\frac{Q^2}{k^2}) \cdot [ x_P G_{BFKL}(x_P,
k^2)]^2
\eeq
The integration over $k^2$ in the above expression leads
to an infrared divergency
as $ k^{2} \rightarrow 0 $.
However, the BFKL Pomeron is associated with an
anomalous dimension $\g(\omega)= \frac{1}{2}$, and thus we get Eq.(1)
for which the BFKL gluon distribution is given by
\beq
x_P G_{BFKL}(x_P,k^2)\,\,=\,\,\frac{G_0}{\sqrt{28 N_c \a_s(Q^2_0)
\zeta(3)}}
\,\,
$$
$$
\cdot\,
\sqrt{\frac{k^2}{Q^2_0}}\,\cdot\,\frac{1}{\sqrt{\ln \frac{1}{x_P}}}\,
\cdot\,
(\frac{1}{x_P})^{\omega_0}\,\cdot\,e^{- \,\frac{\pi(\ln\frac{k^2}{Q^2_0})
^2}{
56 N_c \a_s(Q^2_0)\zeta(3) \ln\frac{1}{x_P}}}
\eeq
Substituting Eq.(8) in Eq.(7), we  rewrite
Eq.(7) in a more
compact form using new variables $y_P\,\,=\,\,\ln\frac{1}{x_P}$,
$y_B\,\,=\,\,\ln\frac{1}{x_B}$, $r_Q\,\,=\,\,\ln \frac{Q^2}{Q^2_0}$ and
$r = \ln \frac{k^2}{Q^2_0}$. This yields
\beq
\Delta F_2 \,\,=\,\,-\,\,\frac{\gamma}{Q^2_0} \,\,\int^{y_B}_0
 d y_P \int^{r_Q}_0 d r
F_2^{GLAP}(\frac{x_B}{x_P},\frac{Q^2}{k^2})\, \cdot \,\frac{\pi}
{\Delta y_P}
\cdot \,e^{ 2 \omega_0 \,y_P - \frac{ 2 r^2}{\Delta y_P}}
\eeq
where $\Delta = 56 \zeta(3) \bar \a_s$ and $\bar \a_s = \frac{N_c \a_s}
{\pi}$.
\newline
\par
We wish to stress that  the infrared divergence of the above integral
should be studied
in more detail.  To this end we consider the situation where
$r_Q$ and $y_B$ are sufficiently large
so that we can use the solution of the
GLAP evolution equation in the region of small $x_B$ to assess
$F_2^{GLAP}$ in Eq.(9). We obtain
\beq
F_2^{GLAP}\,\,=\,\,A\,e^{2 \sqrt{\bar \a_s (y_B - y_P) ( r_Q - r)}}
\eeq
We fix  $\a_s$  so as
to perform our calculations in a way consistent with the BFKL equation.
There is no danger in doing so, as the $Q^2$
variation in the small $x_B$ HERA
kinematic region is negligible, allowing us to
 use this approach for the analysis of the
HERA data. We absorb all irrelevant  factors
appearing before the exponential in a constant factor A, which appears
 in front of the expression.
\newline
\par
Substituting $F_2^{GLAP}$
in Eq.(9) we reduce the equation to the form
\beq
\Delta F_2 \,\,=\,\,-\,\,\frac{\gamma A}{Q^2_0} \,\,\int^{y_B}_0
 d y_P \int^{r_Q}_0 d r
\,\,e^{2 \sqrt{\bar \a_s (y_B - y_P) ( r_Q - r)}}
\,\cdot \,\frac{\pi}{\Delta y_P}
\cdot \,e^{ 2 \omega_0 \,y_P - \frac{ 2 r^2}{\Delta y_P}}
\eeq
It is easy to see that there is no saddle point in the integration with
respect to $r$. Indeed, the equation for the saddle point is
\beq
\frac{\partial \Psi}{\partial r} \,\,=\,\,0
\eeq
where
\beq
\Psi \,\,=\,\,2 \sqrt{\bar \a_s (y_B - y_P) ( r_Q - r)}
\,\,+\,\, 2 \omega_0 \,y_P - \frac{ 2 r^2}{\Delta y_P}
\eeq
Eqs.(12,13) give
\beq
-\,\sqrt{\frac{\bar \a_s (y_B - y_P)}{r_Q - r}} \,\,-\,\,\frac{4r}{\Delta
y_P}
\,\,=\,\,0
\eeq
The saddle point can only be at negative values of
$r$, but one cannot
 trust the
BFKL equation in this domain,  where the virtuality $k^2$ is less
than $Q_0^2$.
At such small values of virtualities there are  certainly  large
corrections, and it does not seem reasonable to expect the BFKL Pomeron
description to be valid in this region
\footnote{The position of the saddle point in the
region of
small virtualities has been studied in all details in Ref.\cite{CCH}.}.
\newline
\par
We note  that the most important region of integration
is
still $r \,\rightarrow \,0$, or in other words, the dominant value of
$k^2$ remains $k^2 \sim Q_0^2$. This leads us to conclude
 that the BFKL contribution is
 questionable and one needs to study the integral of Eq.(11) in more
detail, so as to be sure of the domain where it is valid.
To this end, we observe
that our integral over $y_P$ has a very good saddle
point. Indeed, the equation for this saddle point is
\beq
\frac{\partial \Psi}{\partial y_P} \,\,=\,\,0\,\,=
-\,\sqrt{\frac{\bar \a_s ( r_Q - r)}{y_B - y_P}}  \,+\,2 \omega_0 +
\frac{2 r^2}{\Delta y^2_P}
\eeq
Neglecting the last term, we have the saddle point value for $y_P$
\beq
y^{SP}_P\,\,=\,\,y_B \,-\,\frac{\bar \a_s (r_Q - r)}{ 4 \omega^2_0}
\eeq
 We now  check the reliability of
the GLAP approach for the calculation of $F_2^{GLAP}$.
We recall that the typical value of $\omega$, the argument
of the anomalous dimension of the GLAP equation\footnote{
We denote $\omega$ = N - 1, where N is the moment variable.},
is given by
\beq
\omega\,\, = \,\,\sqrt{\frac{\bar \a_s (r_Q - r)}{y_B - y_P}}
\eeq
Substituting  $y_P \,=\,y^{SP}_P$, we have
$\omega \,\,=\,\,2 \,\omega_0$.
\newline
\par
The BFKL anomalous dimension is given by the series\cite{JAR}
\beq
\gamma( \omega)\,\,=\,\,\frac{\bar \a_s}{\omega}\,\,+\,\,2 \zeta(3) \,
(\frac{\bar \a_s}{\omega})^4 \,\,+\,\,O (\frac{ (\bar \a_s )^5}{\omega^5}
)
\eeq
Substituting
$\omega =  2 \omega_0$, we  see that the BFKL corrections are very
small. This does not mean that we do not need the normal GLAP corrections,
which are essential (see ref.\cite{EKL}), but they cannot change the
main result of the  present
problem.
\newline
\par
Substituting $y_P = y^{SP}_P $ in Eq.(10), we end up with the following
integral over $r$ to be inserted in Eq.(9)
\beq
\int^{r_Q}_0 d r \,\,e^{ 2 \omega_0 y_B \,+\,\frac{\bar \a_s}{2 \omega_0}
\,( r_Q - r)\,-\,\frac{2 r^2}{\Delta y_B}}
\eeq
for
$y_B \,\gg \,\frac{\bar \a_s (r_Q - r)}{ 4 \omega^2_0}$. This is the
kinematic region
which is most interesting both from the theoretical and experimental
points of view.
One can see that the integral over $r$ in Eq.(19) is concentrated at
small $r\,\sim\,\frac{\omega_0}{\bar \a_s} \,\propto\,0(\a_s)$ and
at large $r_Q$ and $y_B$. We face the dilemma of how much trust one can
put on  the perturbative calculation which  estimates  the behaviour of the
deep inelastic gluon distribution, with virtualities of the
order of $Q^2_0$. Apparently, this calculation is not reliable as the problem
reduces to that of the
 energy behaviour of the typical hadron - hadron
interaction
at high energy, which is described by the "soft"
Pomeron\cite{DL}\cite{GLM3}.
\newline
\par
Despite this reservation,
the above statement is certainly correct if we consider
only small
$\a_s$ and/or large $ Q^2 $,  the $\ln (1/x_B)$ parameters in our
calculation.
Numerically,
the situation is more promising. Indeed, as we shall show,
the HERA data on $F_2$ \cite{ZEUS}\cite{H1}
confirm the theoretical expectation that the
BFKL Pomeron with $\omega_0$ = 0.5 contributes at low $x_B$. Substituting
$\omega_0$ = 0.5 we obtain a typical value of
$r\,\simeq\,\frac{2 \omega_0}{\bar \a_s}\,\approx 6$, in the integral
of Eq.(19). This value is sufficiently large to justify
our using pQCD, to evaluate the diagram of Fig.2. Moreover,
$r \,\approx\,$ 6 is larger than the value of $r_Q$ in HERA kinematic
region, so we can estimate the value of the integral in Eq.(19)
as $r_Q$.
Collecting all factors together, we obtain  Eq.(4), which was used in our
description of the HERA data.
\newline
\par
Detailed comparisons of our calculations with the
data\cite{ZEUS}\cite{H1}
are displayed in Fig.3 where we present $F_2(W,Q^2)$ and
$F_2(x_{B},Q^2)$.
We did not attempt a "best fit",
nevertheless, our ability to reproduce the gross features of the data is
 evident.
The following comments relate to the
data choice and detailed features of our fit:
\\
1) The data considered are bounded by $x_{B} \leq 10^{-2}$ and
$W \geq 50 \,GeV$.
\\
2) The following parameters were used in the numerical fit:
$\tilde G_0$ = 0.024, \\ $\gamma$ = 0.015 $GeV^2$, $Q_0$ = 1 $GeV$ and
C = -0.025.
With these parameters we get that
$\frac{F^D_2}{F_2} \,\approx 0.30$ at $Q^2$ = 8.5 $GeV^2$
and $W \approx $ 250 $GeV$.
We can reproduce the data with a
$\frac{F_2^D}{F_2}$ which is smaller, but, clearly, our
requirement for a relatively large DIS diffractive component is essential
for this approach.
\\
3) As can be readily seen from Fig.3, we obtain a reasonable
 description of the data
down to  values  of $Q^2 \approx 3\, GeV^2$. At smaller values of $Q^2$
we require larger  SC  to reproduce the data.
\newline
\par
Our inability to reproduce the lowest $Q^2$ data is not surprising.
Clearly, the present approach is over simplified,  as we fail to take
into account the different effects of the SC on the diffractive
and the total DIS cross sections. We call attention to the
observation\cite{GLM2}\cite{GLM3} that in soft hadron interactions
$\sigma_{diff}(with \,SC) \approx 0.3 \sigma_{diff}(without \,SC)$
whereas
$\sigma_{t}(with \,SC) \approx 0.75 \sigma_{t}(without \,SC)$. As we have
shown\cite{GLM4} the strength of the SC in pQCD is determined by a
parameter $\kappa \,=\, \frac{3 \alpha_{s} \pi}{2 B Q^2}[xG(x,Q^2)]$,
where B is the elastic slope.
At
low $Q^2$, $\kappa \approx 1$, and  we doubt the validity of Eq.(5),
which is at the core of our calculation. Nevertheless, even when
$\kappa\,\approx 1$, one can still use Eq.(3) with $\Delta F_2$ as
defined in Eq.(4), and obtain a better assessment of $F_2$ at low $Q^2$
than the one we have presented here.
The reason for this is, that the SC
to the total DIS cross section turns out
to be smaller than
the SC to diffraction dissociation.
This problem is
connected to the broader issue of the transition between the (hard)
pQCD and the (soft) non perturbative domain, which we plan to discuss in
a forthcoming publication.
\newline
\newline
{\bf ACKNOWLEDGEMENTS:}
E. Levin and U. Maor acknowledge the financial support by CNPq and
the hospitality extended to them by LAFEX (CBPF). E. Gotsman
would like to thank his colleagues at Irvine for their warm hospitality.
\newline
\newline
\section*{Figure Captions}
{\bf Fig.1:} $W$ dependance of $F_2(W,Q^2)$. Data is taken from
ZEUS\cite{ZEUS} and H1\cite{H1} with $Q^2 \geq 8.5 \,GeV^2$.\\
v\newline
{\bf Fig.2:} Diffraction dissociation in perturbative QCD.\\
\newline
{\bf Fig.3:} Comparison of $F_2(W,Q^2)$ and $F_2(x,Q^2)$
data with our calculations.\\
\newline


 
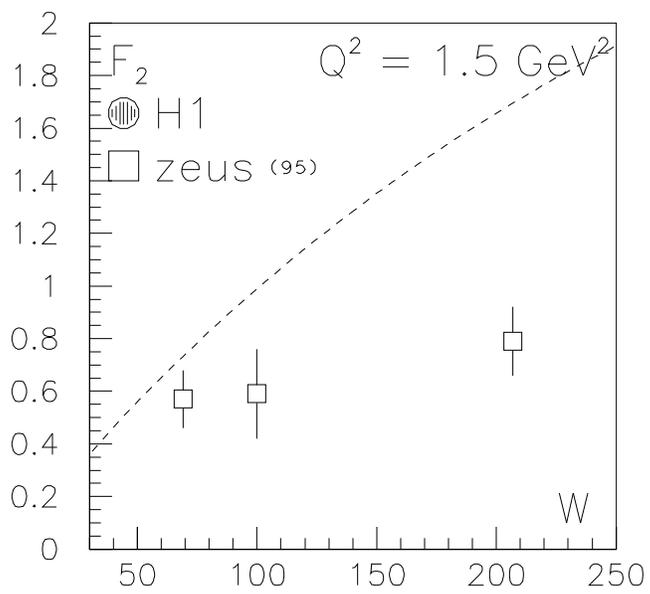
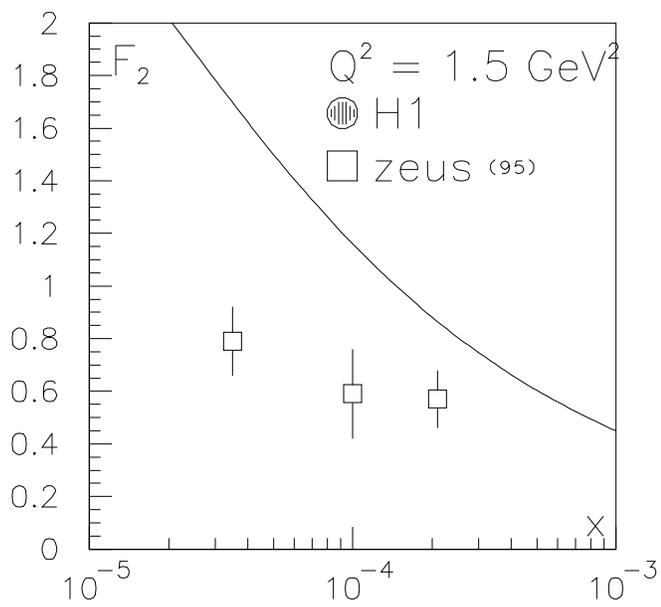
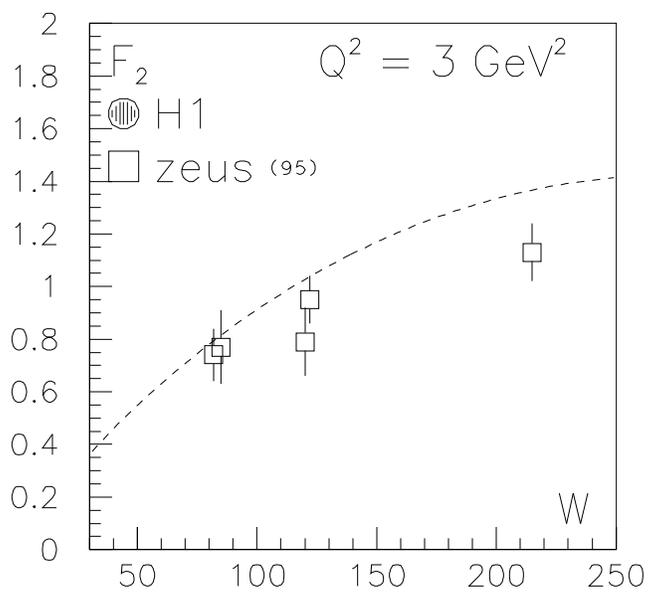
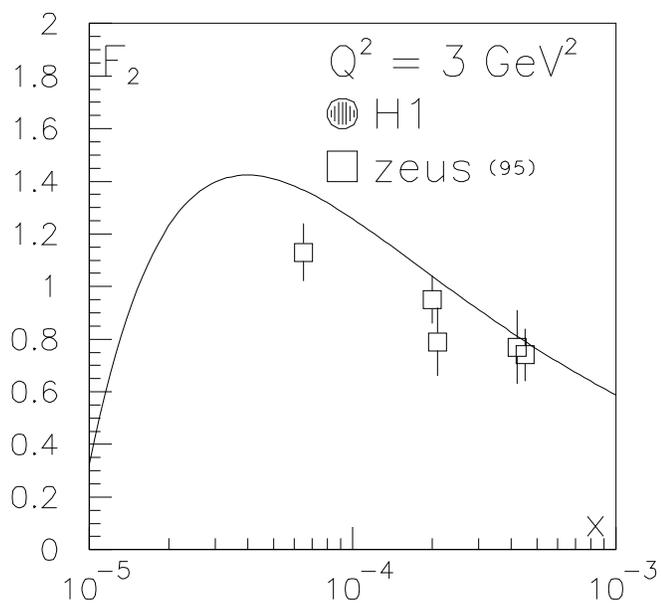
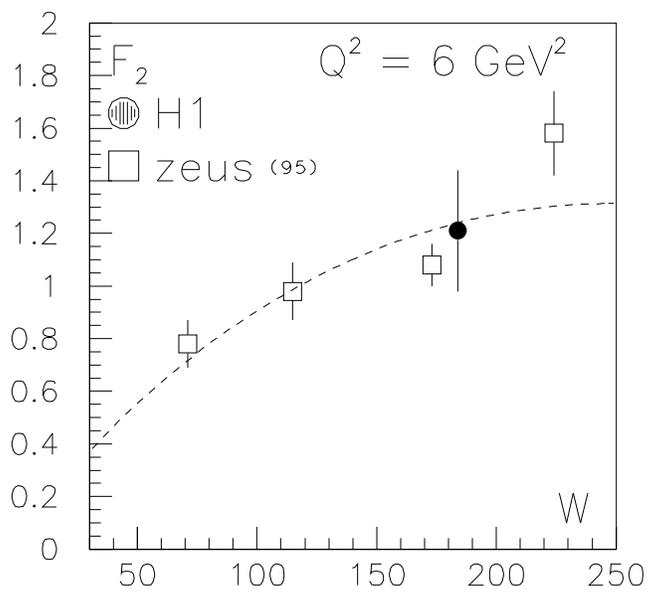
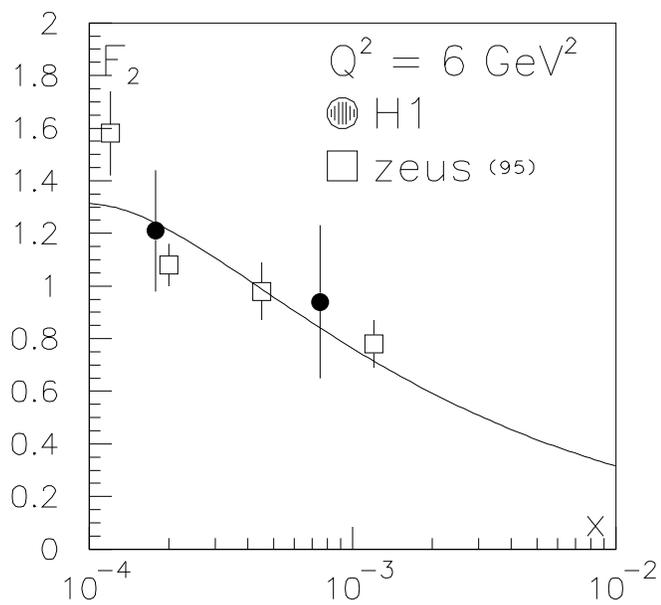
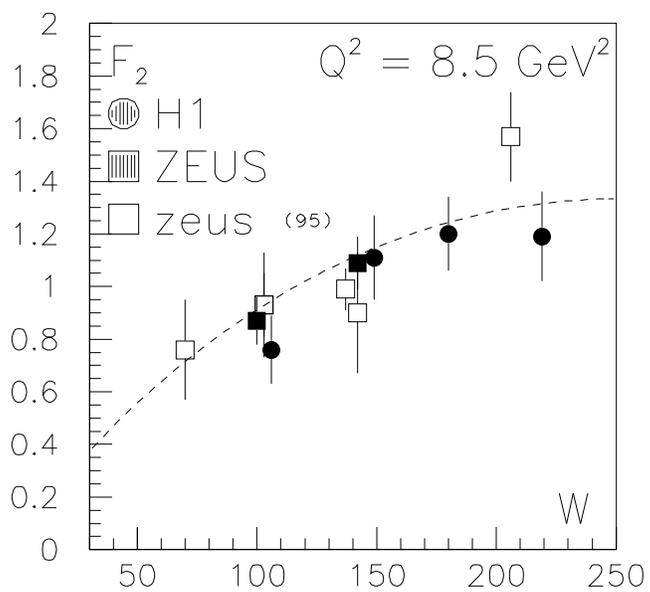
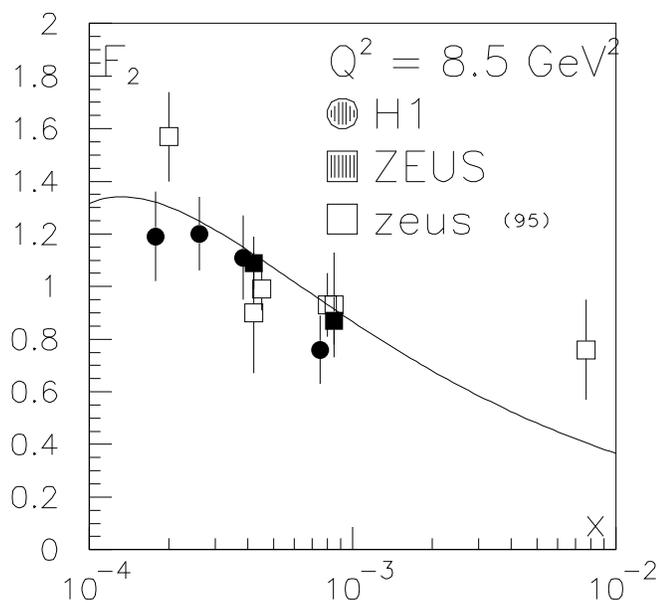
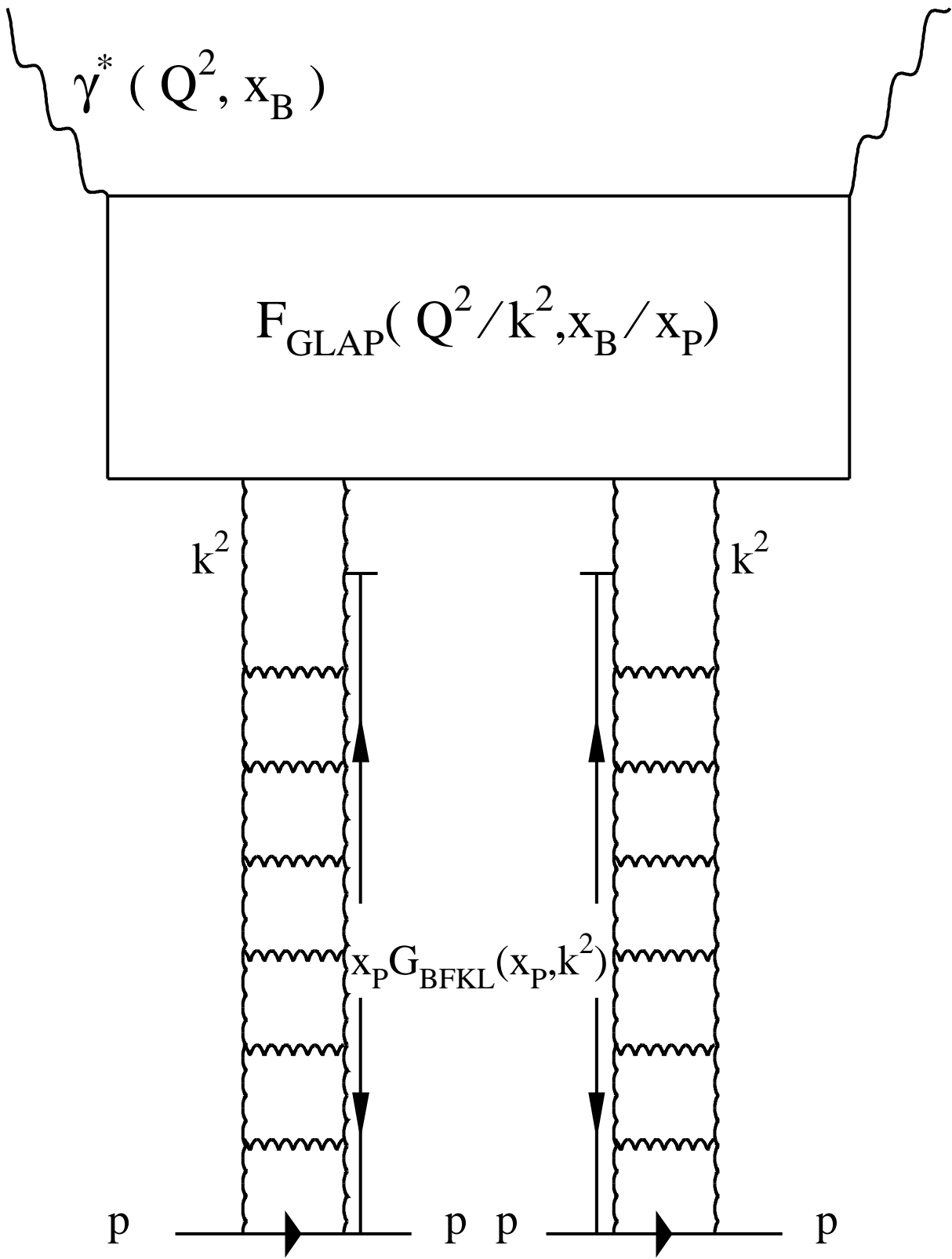
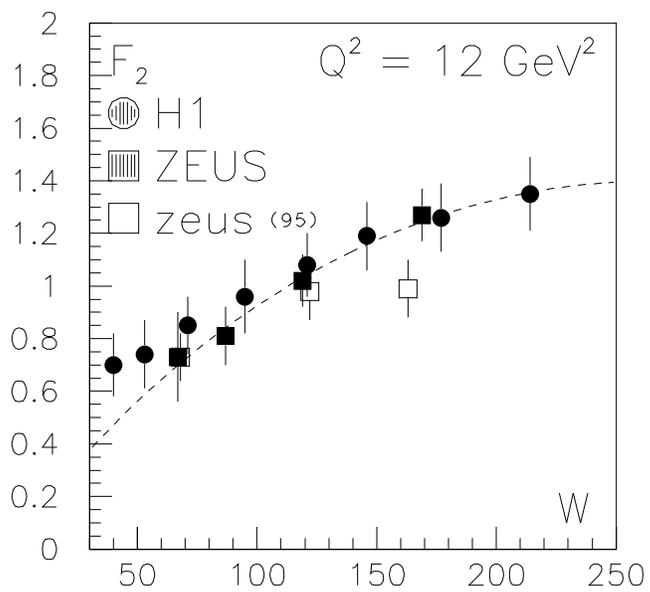
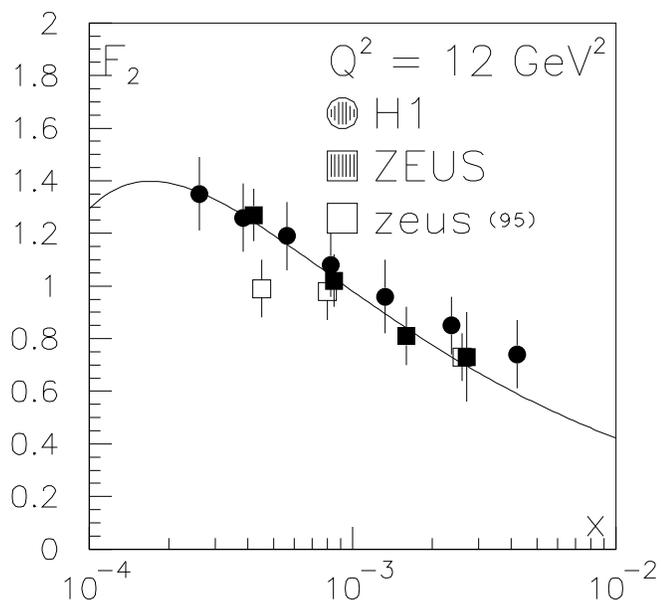
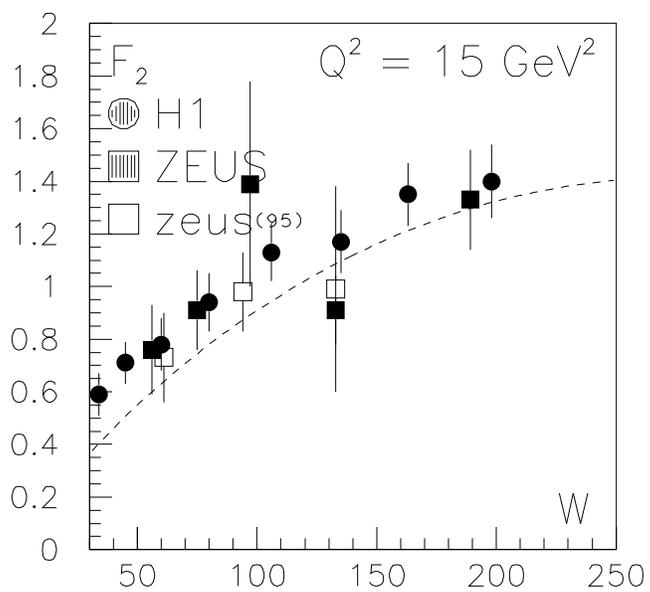
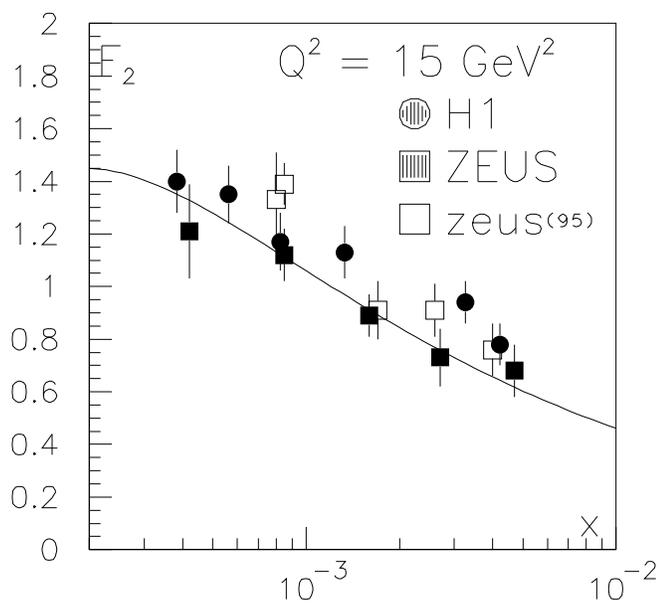
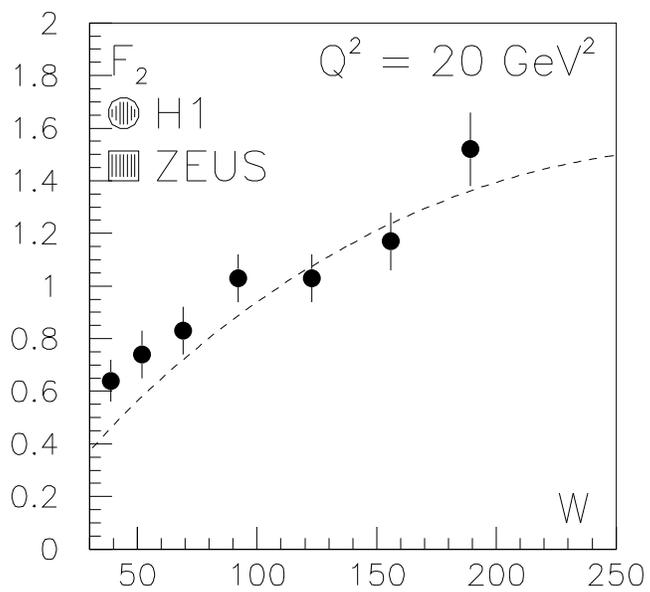
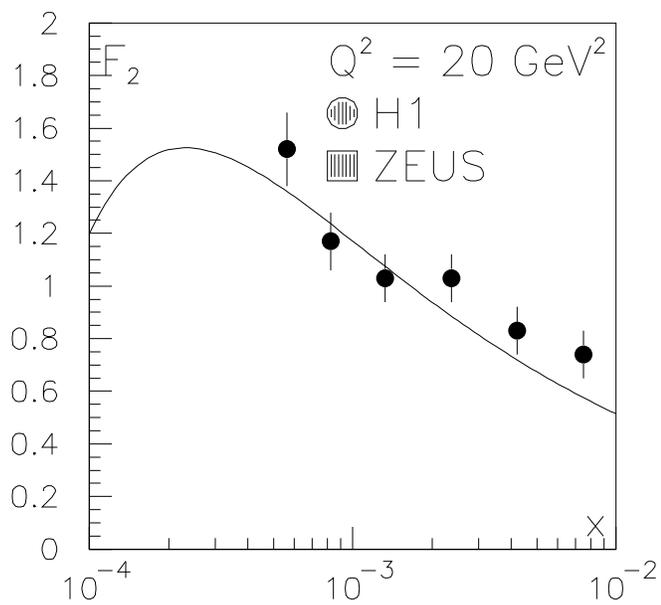
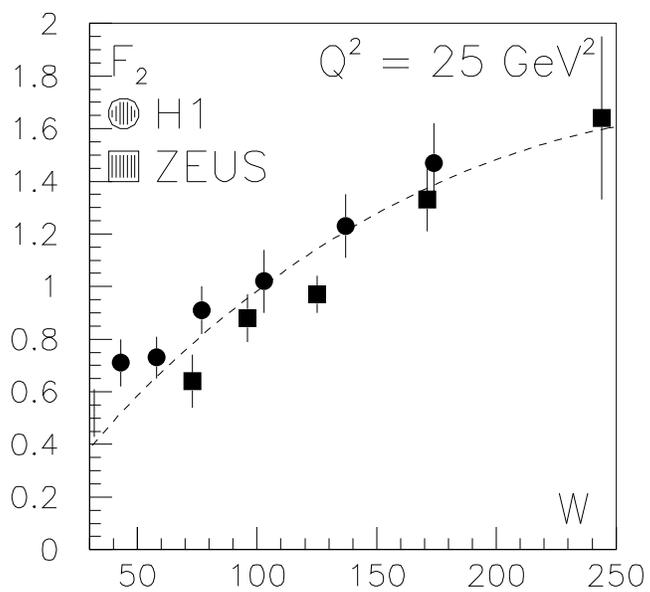
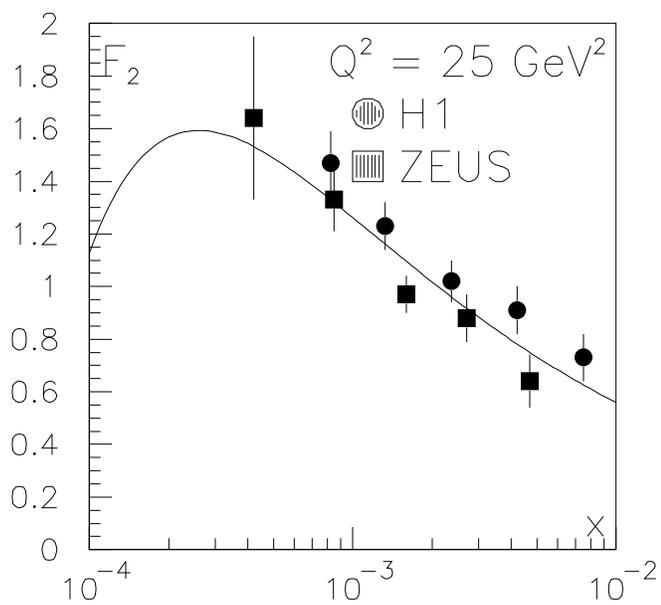
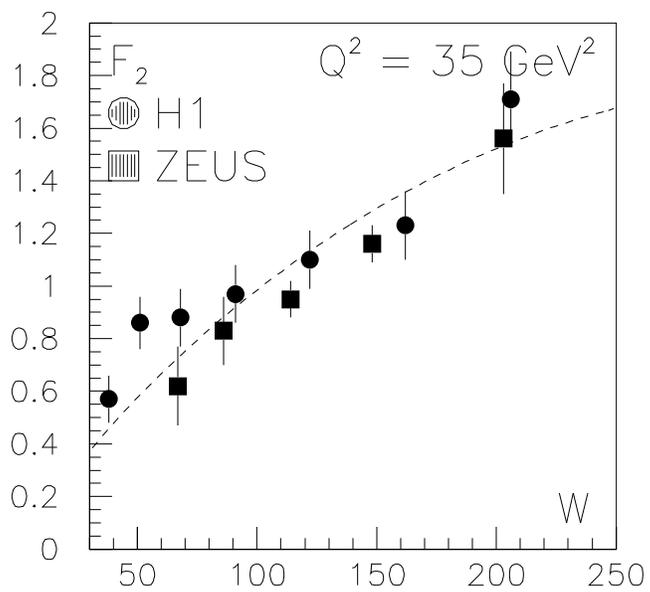
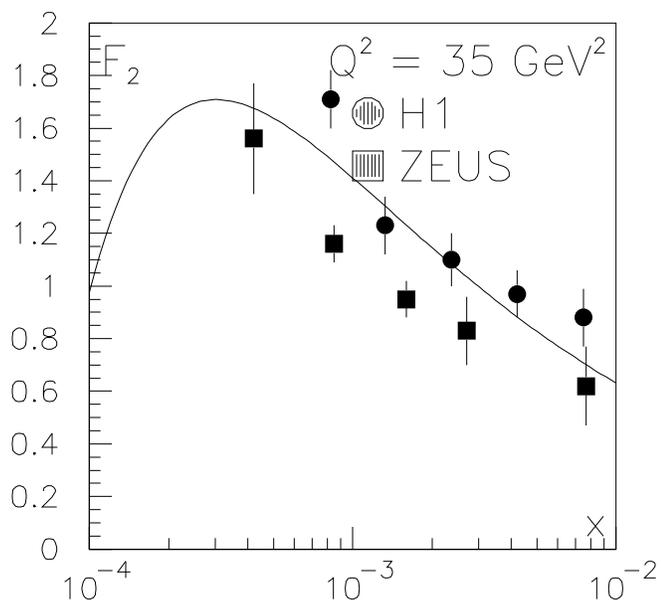
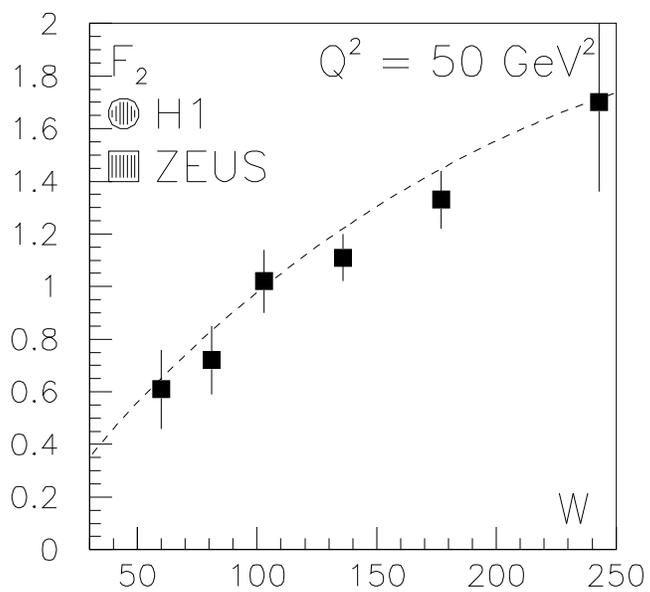
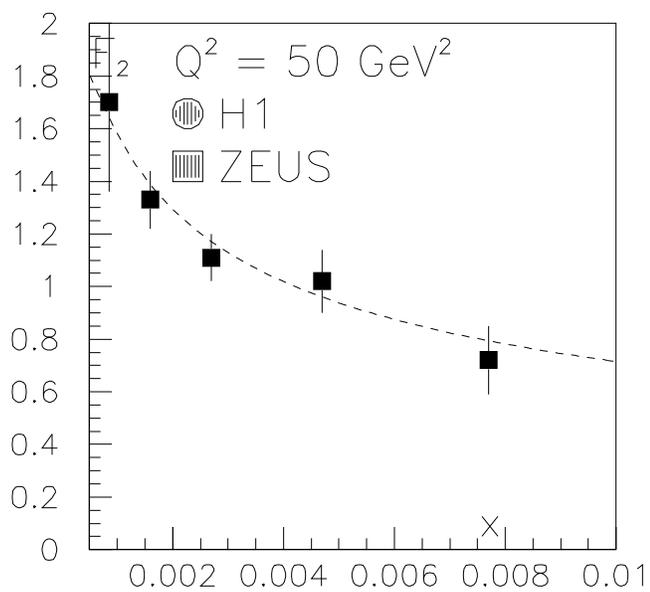
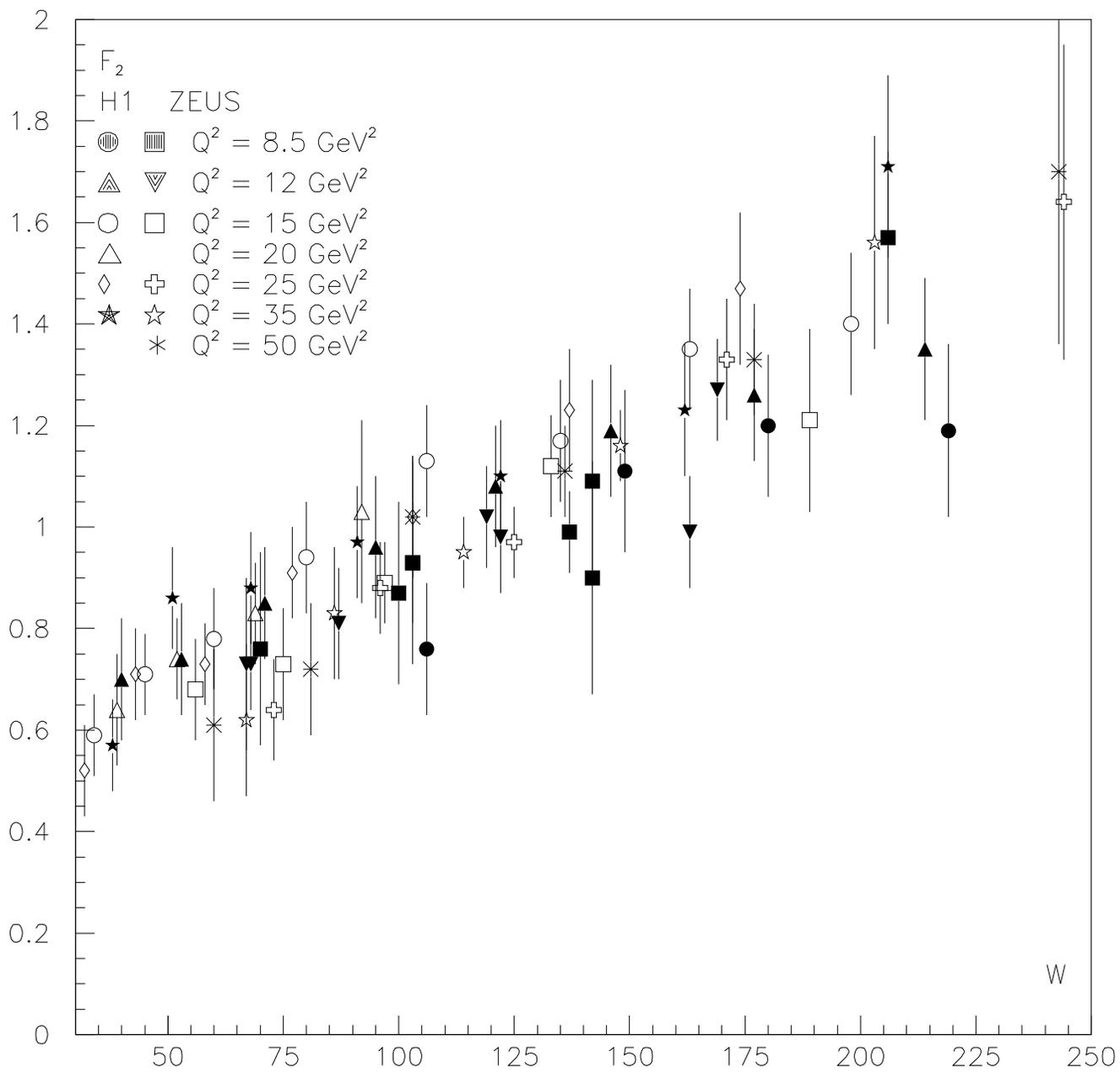
\end{document}